\documentclass{article}
\usepackage{kmn,latexsym,pifont,graphics}
\newif\ifAMStwofonts

\newcommand{\gtsim}{\mbox{{\raisebox{-0.4ex}{$\stackrel{\star}
{{\phantom{o}}} $}}}}

\def\source{LMC~X--2}
\def\til{$\sim$}

\def\deg{$^{\circ}$}

\def\lx{$L_{x}$}

\def\leq{\hbox{${_<\atop{\sim}}$}}
\def\geq{\hbox{${_>\atop{\sim}}$}}
\def\ang{\thinspace\hbox{\AA}}
\def\ergsec{\thinspace\hbox{$\hbox{erg}\thinspace\hbox{s}^{-1}$}}

\def\arcsecdot{\nobreak\ifmmode{''\hskip-0.45em.\hskip0.08em}%
                         \else{$''\hskip-0.45em.\hskip0.08em$}\fi}
\def\arcsec{\nobreak\ifmmode{''\hskip-0.45em}%
                      \else{$''\hskip-0.45em$}\fi}

\title[Searching for periodicities in LMC~X--2]
      {Searching for periodicities in the MACHO light curve of LMC~X--2}
\author[C. Alcock et al.]
{C. Alcock$^{1,2}$, R. A. Allsman$^{3}$, D. R. Alves$^{4}$, 
T. S. Axelrod$^{5}$, A. C. Becker$^{6}$, 
\and
D. P. Bennett$^{7}$, P. A. Charles$^{8}$, K. H. Cook$^{1,2}$\thanks{Visiting 
Astronomer, Cerro Tololo Inter-American Observatory, which is operated by the 
Association of Universities for Research in Astronomy, Inc., under cooperative 
agreement with the National Science Foundation.}, A. J. Drake$^{1}$,  
K. C. Freeman$^{5}$, 
\and
M. Geha$^{9}$, K. Griest$^{2,10}$, P. Groot$^{11}$\thanks{Present address: 
Harvard-Smithsonian Center for Astrophysics}, M. J. Lehner$^{12}$, 
S. L. Marshall$^{1}$\gtsim,
\and
K. E. McGowan$^{8}$\thanks{Email: kem@astro.ox.ac.uk},  D. Minniti$^{1,13}$,
C. A. Nelson$^{14}$, B. A. Peterson$^{5}$, P. Popowski$^{1}$, 
\and M. R. Pratt$^{15}$, P.J. Quinn$^{16}$, W. Sutherland$^{8}$, 
A. B. Tomaney$^{6}$, T. Vandehei$^{10}$, 
\and
J. van Paradijs$^{11,17}$\\
$^{1}$Lawrence Livermore National Laboratory, Livermore, CA 94550, USA\\
$^{2}$Center for Particle Astrophysics, University of California, Berkeley, CA 
94720, USA\\
$^{3}$Supercomputing Facility, Australian National University, Canberra, ACT 
0200, Australia\\
$^{4}$Space Telescope Science Institute, Baltimore, MD 21218, USA\\
$^{5}$Research School of Astronomy and Astrophysics, Canberra, Weston Creek, 
ACT 2611, Australia\\ 
$^{6}$Departments of Astronomy and Physics, University of Washington, Seattle, 
WA 98195, USA\\
$^{7}$Department of Physics, University of Notre Dame, Notre Dame, IN 46556, 
USA\\
$^{8}$Department of Astrophysics, Nuclear Physics Laboratory, Keble Road, 
Oxford OX1 3RH\\
$^{9}$ Department of Astronomy and Astrophysics, UC Santa Cruz, Santa Cruz, CA
90064\\
$^{10}$Department of Physics, University of California, San Diego, La Jolla, CA
92093, USA\\
$^{11}$Astronomical Institute `Anton Pannekoek'/CHEAF, University of Amsterdam,
Kruislaan 403, 1098 SJ Amsterdam, The Netherlands\\
$^{12}$Department of Physics, University of Sheffield, Sheffield, S3 7RH\\
$^{13}$Departmento de Astronomia, P. Universidad Catolica, Casilla 104, 
Santiago 22, Chile\\
$^{14}$Department of Physics, University of California, Berkeley, CA 94720, 
USA\\
$^{15}$Center for Space Research, MIT, Cambridge, MA 02139, USA\\
$^{16}$European Southern Observatory, Karl-Schwarzchild Strasse 2, D-85748, 
Garching, Germany\\
$^{17}$Physics Department, University of Alabama in Hunstville, Huntsville, AL 
35899, USA}

\date{Accepted 
Received}

\pagerange{\pageref{firstpage}--\pageref{lastpage}}
\pubyear{2000}

\begin{document}

\maketitle

\label{firstpage}

\begin{abstract}
{Using the exceptional long-term monitoring capabilities of the MACHO project, 
we present here the optical history of \source\ for a continuous 6-yr period.  
These data were used to investigate the previously claimed periodicities for 
this source of 8.15 h and 12.54 d : we find upper amplitude limits of 0.10 mag
and 0.09 mag, respectively.}

\end{abstract}


\section{Introduction}

\source\ was discovered in the Large Magellanic Cloud (LMC) by early satellite 
flights (Leong et al.\ 1971) which showed it to be a variable X-ray source with
\lx\til(0.6--3)x10$^{38}$\ergsec\ (Markert \& Clark 1975; Johnston, Bradt \& 
Doxsey 1979; Long, Helfand \& Grabelsky 1981).  \source\ is the most luminous 
low mass X-ray binary (LMXB) known. After its precise position was determined 
(Johnston et al.\ 1979), the optical counterpart was identified as a faint, 
{\it V} \til18.8, blue star similar to Sco~X--1 (Pakull 1978; Pakull \& Swings 
1979).

Motch et al.\ (1985) found variations of \til0.4 mag in the optical light curve
of \source, which were modulated on a period of 6.4 h (see also Bonnet-Bidaud 
et al.\ 1989).  More extensive photometric observations obtained by Callanan et
al.\ (1990) showed flickering of \til0.1 mag on timescales of \leq0.5 h and
\til0.3 mag variations on \til4 h timescales.  Period searching revealed a 
period of 8.15$^{+}_{-}$0.02 h with a semi-amplitude of \til0.08 mag which, 
being stable over their \til14 d observations, they interpreted as being 
orbital in origin.  However, Crampton et al.\ (1990) suggested a much longer 
orbital period of 12.54 d with a semi-amplitude of \til0.5 mag from their 
photometric observations that spanned nine nights.  This substantial 
disagreement over LMC X-2's fundamental parameter, its orbital period, has 
still not been resolved and is, of course, crucial to any detailed 
interpretation of the source.

\subsection{Is the Orbital Period 8.15 h?}

If an orbital period of 8.15 h is correct, we would expect a significant 
contribution from a heated secondary or outer disc bulge, and this should be 
evident from the shape of the folded light curve (Callanan et al.\ 1990).  
However, as Crampton et al.\ (1990) did not detect the 8.15-h modulation, it 
implies that the amplitude of this modulation is itself variable on timescales 
\geq10 days.  If it is owing to X-ray heating of the secondary, a 
semi-amplitude of 0.08 mag indicates that i\leq70\deg\ (van Paradijs, 
van der Klis \& Pederson 1988).

\subsection{Or is it 12.54 d?}

If a 12.54-d period is correct, taking into account the structure and asymmetry
of the light curve obtained, \source\ would be similar to Cyg X-2 (Cowley, 
Crampton \& Hutchings 1979).  This would suggest that the variations in the 
\source\ light curve might contain both a heated component and a partial 
eclipse of the accretion disc by the secondary.  Irregularities in the disc 
cause some of the pre-eclipse modulation that gives rise to pre-eclipse dips 
and variability as observed in other LMXBs (see White 1989).  In a long-period 
LMXB, only an evolved star can fill the Roche lobe, and the high X-ray 
luminosity of \source\ may be due to the high rate of mass transfer on to the 
neutron star as the companion evolves along the giant branch (Webbink, 
Rappaport \& Savonije 1983).  The period found could also be as a result of 
disc precession rather than orbital variations, or a beat between a photometric
and orbital period (Crampton et al.\ 1990).  Short periodicities must also be 
present for this to be true, and although none were found by Crampton et al.\ 
(1990), the periods found by Callanan et al.\ (1990) and Bonnet-Bidaud et al.\ 
(1989) may therefore be real.

Evidence for the long period would be the direct detection of the secondary in 
the spectrum (cf. Cyg X--2; Casares, Charles \& Kuulkers 1998 and references 
therein), but thus far the optical spectra show no absorption features that can
be attributed to the companion star (Bonnet-Bidaud et al.\ 1989).  Lack of 
detection could be as a result of the extreme brightness of the accretion disc.
This may be owing to the high luminosity fuelled by the higher rate of mass 
transfer allowed by the lower metal abundances in the LMC, leading to a higher 
Eddington luminosity (Russell \& Dopita 1990).  No correlation between X-ray 
and optical light curves has yet been detected that would help in the 
identification of the period.

We present here the results of \til6 years of optical monitoring of the source.
These observations were acquired as a by-product of the MACHO project (Alcock 
et al.\ 1995a), owing to the serendipitous location of \source\ in a surveyed 
field.  Such extended monitoring is ideal for investigating modulations on 
timescales of tens of days.

\section{Observations}

The MACHO observations were made using the 1.27-m telescope at Mount Stromlo 
Observatory, Australia.  A dichroic beamsplitter and filters provide 
simultaneous CCD photometry in two passbands, a 'red' band (\til6300--7600 
\ang) and a 'blue' band (\til4500--6300 \ang).  The latter filter is a broader
version of the Johnson {\it V} passband (see Alcock et al.\ 1995a, 1999 for 
further details).

The images were reduced with the standard MACHO photometry code {\scshape 
sodophot}, based on point-spread function fitting and differential photometry 
relative to bright neighbouring stars.  Further details of the instrumental 
set-up and data processing may be found in Alcock et al.\ (1995b, 1999), 
Marshall (1994) and Stubbs et al.\ (1993).

\section{Results}

Using the absolute calibration of the MACHO fields the red and blue magnitudes 
were transformed to Johnson {\it V} and Cousins {\it R} passbands (Fig. 1). 
The data consist of MACHO project observations from 1992 November 1 to 1998 
November 21.  The sampling of the light curves increases at \til1800 d, when 
the MACHO observing strategy was modified to give increased weight to fields 
further from the LMC bar.  This epoch coincides with the increase in the mean 
magnitude of \source.

To search for periodicities, after the long-term trends in the {\it V}- and 
{\it R}-band datasets were removed by subtracting a third order polynomial fit,
two different frequency domain techniques were employed : (i) we calculated a 
Lomb-Scargle (LS) periodogram (Lomb 1976; Scargle 1982) on each dataset, to 
search for sinusoidal modulations [this periodogram is a modified discrete 
Fourier transform (DFT), with normalizations that are explicitly constructed 
for the general case of time sampling, including uneven sampling; see Scargle 
1982]; (ii) we constructed a phase dispersion minimisation periodogram (PDM), 
which works well even for highly non-sinusoidal light curves (see Stellingwerf 
1978).  

As there are two candidate orbital periods for \source\ (\til8.2 h or 
\til12.5 d), searches were made with optimal ranges and resolution for each 
period.  For the shorter period a frequency space of 1-10 cycle d$^{-1}$ was 
searched with a resolution of 0.0001 cycle d$^{-1}$, and for the longer period 
a frequency space of 0.01-1 cycle d$^{-1}$ was searched with the same 
resolution as above.  No significant dips were found in the PDM plots for 
either {\it V}- or {\it R}-band.  The short periodicity  LS periodogram for the
{\it V}-band data shows one peak that is just above the 99\% confidence limit 
(Fig. 2).  This peak corresponds to 2.68 h.  However, no significant peaks 
were found in the LS for the {\it R}-band data.  Searches were also carried out
on the better sampled section of {\it V}- and {\it R}-band data, with no 
significant periodicity being found with either method.

By constructing the cumulative probability distribution (CDF) of the random 
variable $P_{\it X}$($\omega$), the power at a given frequency, where {\it X} 
is pure noise (Scargle 1982), we can measure the significance of the peaks in 
the LS periodogram.  A peak power that is above the 99 per cent confidence 
level read from the CDF for the dataset would indicate that a significant 
period had been found.  In practice the CDF was constructed using a 
Monte Carlo simulation method.  Noise sets with the same sampling as the MACHO 
data were generated, and the LS periodogram was run upon each one.  The peak 
power occurring in the periodogram due purely to noise was then recorded.  This
was repeated for ten-thousand noise sets, to produce good statistics.  From 
these values the probability of obtaining a given peak power from pure noise 
can then be calculated and the CDF derived.  In order to test the significance 
of peaks from a given data set, the generated noise sets should have the same 
variance.  Therefore, each noise set was generated from a random number 
generator that takes values from a Gaussian distribution with the same 
variance as the data set.

The {\it V}- and {\it R}-band data were folded on a value found from the 
highest peak in the LS periodogram that was within the 3-$\sigma$ measurement 
error of the \til8.2 h period found by Callanan et al.\ (1990).  The folded 
light curves were binned, so as to search for evidence of any weak underlying 
modulation in the light curve (Fig. 3), but no significant modulation is 
evident.  The datasets were then folded and binned using the previously 
proposed 12.54-d period (Crampton et al.\ 1990), the resultant light curves 
were flat within the errors.  Finally the {\it V}- and {\it R}-band data were 
bin-folded on the most significant peak in the PDM in the region of 12.54 d 
(note that Crampton et al.\ 1990 quote no error for their period and their 
observations only span \til9 d, so we assumed that a significant peak at 
14.06 d is relevant), but again no periodic modulation is found (Fig. 3).

To set limits for the detection of each of the candidate periods in the 
{\it V}- and {\it R}-band datasets, the above CDF technique was repeated using 
the real data as the input to the Monte-Carlo simulations, but also adding a 
sinusoidal signal with given semi-amplitude and period.  The CDF was calculated
using an appropriate range and resolution for both the 8.15 h and the 12.54 d 
periods.  The period for the sinusoid was fixed at the quoted value and the CDF
calculation was repeated for varying values of the amplitude. 

The upper limit for detection of a 12.5 d periodicity in the {\it V}-band data 
at 99 per cent confidence is 9.9 per cent flux modulation semi-amplitude, and 
in the {\it R}-band data is 8.1 per cent.  For an 8.2 h periodicity, the upper 
limit for detection at 99 per cent confidence in the {\it V}-band data is 
11.1 per cent flux modulation semi-amplitude, and 9.3 per cent for the 
{\it R}-band data.

The failure to find a period around 8.2 h may be due to the sampling of the 
MACHO data, which was not optimised for finding short periodicities at the low
semi-amplitude found previously (note the higher noise levels in Fig. 2 ({\it
top left}) and ({\it bottom left}) compared to Fig. 2 ({\it top right}) and 
({\it bottom right})).

The 12.54-d modulation was previously found with a semi-amplitude of 0.5 mag, 
a factor of 5 above our detection limit; if it was present in the MACHO data it
should have been found.

\section{ESO {\it V}-Band Light Curve}

\source\ was also observed with the ESO 0.91-m Danish telescope at La Silla 
between 1997 November 17 - December 12 in order to further investigate the 
12.54-d periodicity found by Crampton et al.\ (1990).  Using a Tek CCD, 
{\it V}-band photometry was obtained with integration times of 300 s.  
Individual CCD frames were trimmed, corrected for bias, and flattened using the
{\scshape iraf} reduction routine {\scshape ccdproc}.  Point spread function 
(PSF) fitting was performed with the {\scshape iraf} implementation of 
{\scshape daophot ii} (Stetson 1987).  Magnitudes of \source\ were determined 
using two standard stars of similar colour.  

The light curve obtained (Fig. 4) shows that the variation of \til0.5 mag 
seen by Crampton et al.\ (1990) is present at our day 7, but it is not repeated
12.54 d later.  A dip of \til0.25 mag also occurs at day 13.5, and the 
source is declining at day 25.  This suggests that the dipping is intrinsic to
the source and the variations are not modulated on a 12.54-d period.  The 
source appears brighter compared to the observations taken by Crampton et al.\ 
(1990), but this may be due to an overall brightening of the source as is also 
seen in the MACHO light curve after day 1800 (Fig. 1).

\section{Discussion}

The results from the MACHO data and the ESO light curve indicate that the 
orbital period of \source\ is not 12.54-d.  There is however evidence of 
non-periodic variations on time-scales of \til tens of days which would explain
the results found by Crampton et al.\ (1990).  Long term variations occur in 
other LMXBs (cf Cyg X-2; Smale \& Lochner 1992; Wijnands, Kuulkers \& Smale 
1996) which are not associated with the orbital period of the system.  In high 
mass X-ray binaries (HMXB) {\it superorbital} periods are known for several 
sources.  Theories to explain these long term variations include precession of 
a tilted accretion disc, precession of the neutron star, mass transfer feedback
and triple systems (see Priedhorsky \& Holt 1987 and Schwarzenberg-Czerny 
1992).  In LMXBs such {\it superorbital} periods are much less common.  It 
is thought that they may be due to radiation-driven warped accretion discs 
(e.g. Wijers \& Pringle 1999) or a disc instability in the system (Priedhorsky
\& Holt 1987).

The light curves from the MACHO and ESO data do not confirm the 8.2-h period 
for \source\ found by Callanan et al.\ (1990).  However, had this periodicity 
been present at the Callanan et al.\ (1990) amplitude we could not have 
detected it (as it is below our upper limit) and so this modulation still 
awaits confirmation.

\section{Acknowledgements}

We thank Celeste Ponsioen for the ESO observations. KEM acknowledges the 
support of a PPARC studentship.  DM is supported by Fondecyt 1990440.  This 
work was performed under the auspices of the U.S. Department of Energy by 
University of California Lawrence Livermore National Laboratory under contract 
No. W-7405-Eng-48.

\newpage

\begin{figure*}
\resizebox{1.0\textwidth}{!}{\rotatebox{-90}{\includegraphics{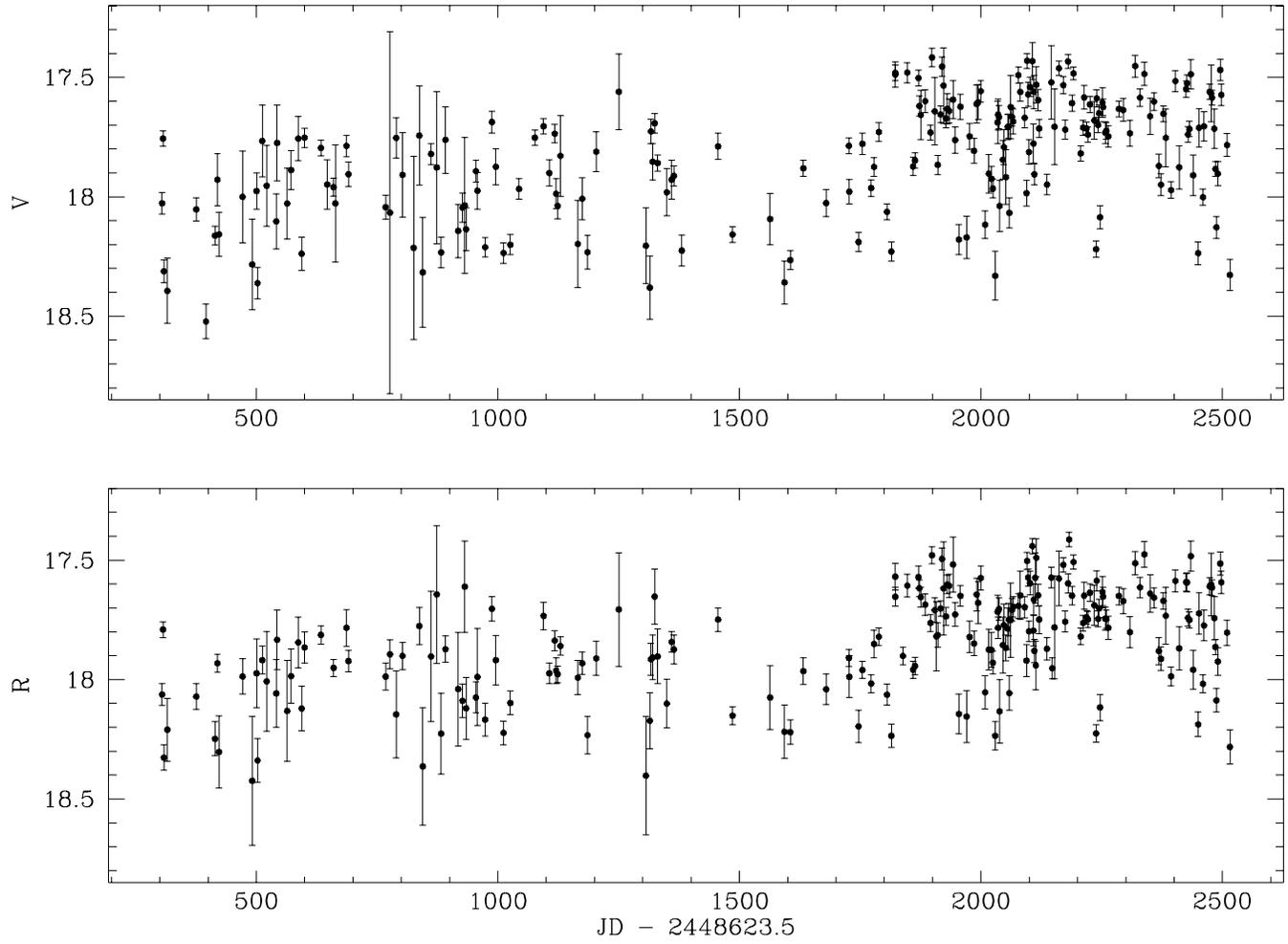}}}
\caption{The optical light curves in blue ({\it top}) and red filters 
({\it bottom}) of \source\ from MACHO project observations.  The {\it V} and 
{\it R} magnitudes have been calculated using the absolute calibrations of the
MACHO fields.  Time zero corresponds to 02 Jan 1992 UT 00:00.}
\end{figure*}

\begin{figure*}
\resizebox{1.0\textwidth}{!}{\rotatebox{90}{\includegraphics{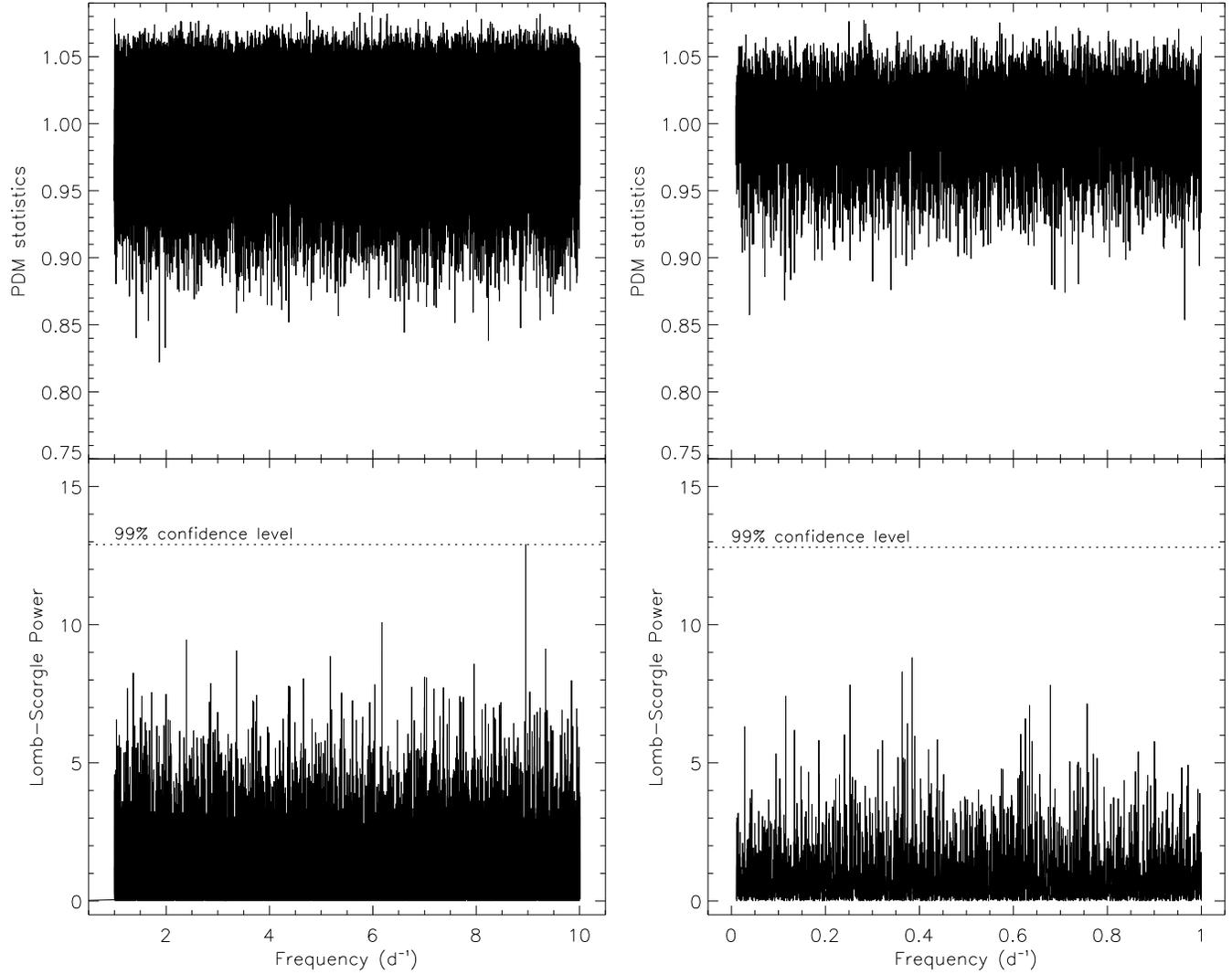}}}
\caption{Results of applying different period searching techniques to our 
\source\ MACHO {\it V}-band light curve.  A phase dispersion minimisation 
analysis of shorter periods (frequency space of 1-10 cycle d$^{-1}$) 
({\it top left}) and longer periods (frequency space of 0.01-1 cycle d$^{-1}$) 
({\it top right}) were searched with a resolution of 0.0001 cycle d$^{-1}$.  
Lomb-Scargle periodogram for shorter periods ({\it bottom left}) and longer 
periods (frequency space and resolution as above) ({\it bottom right}).}
\end{figure*}

\begin{figure*}
\resizebox{1.0\textwidth}{!}{\rotatebox{-90}{\includegraphics{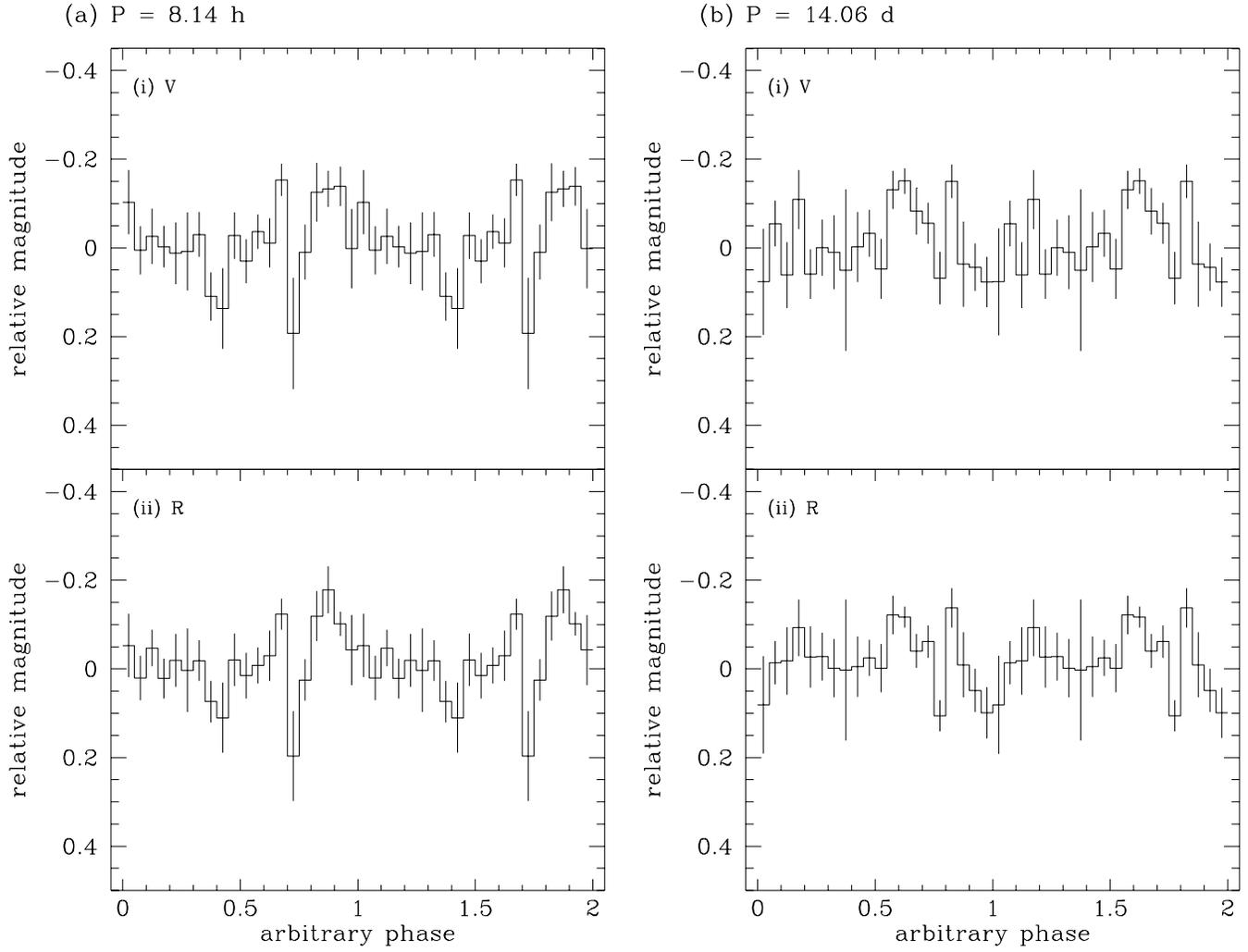}}}
\caption{MACHO photometry of \source\ folded in 20 phase bins on (a) 8.14 h (b)
14.06 d for both {\it V}-band (i) and {\it R}-band (ii).  Error bars for all 
binned light curves are the standard errors for the data points in each bin.}
\end{figure*}

\begin{figure*}
\resizebox{1.0\textwidth}{!}{\rotatebox{-90}{\includegraphics{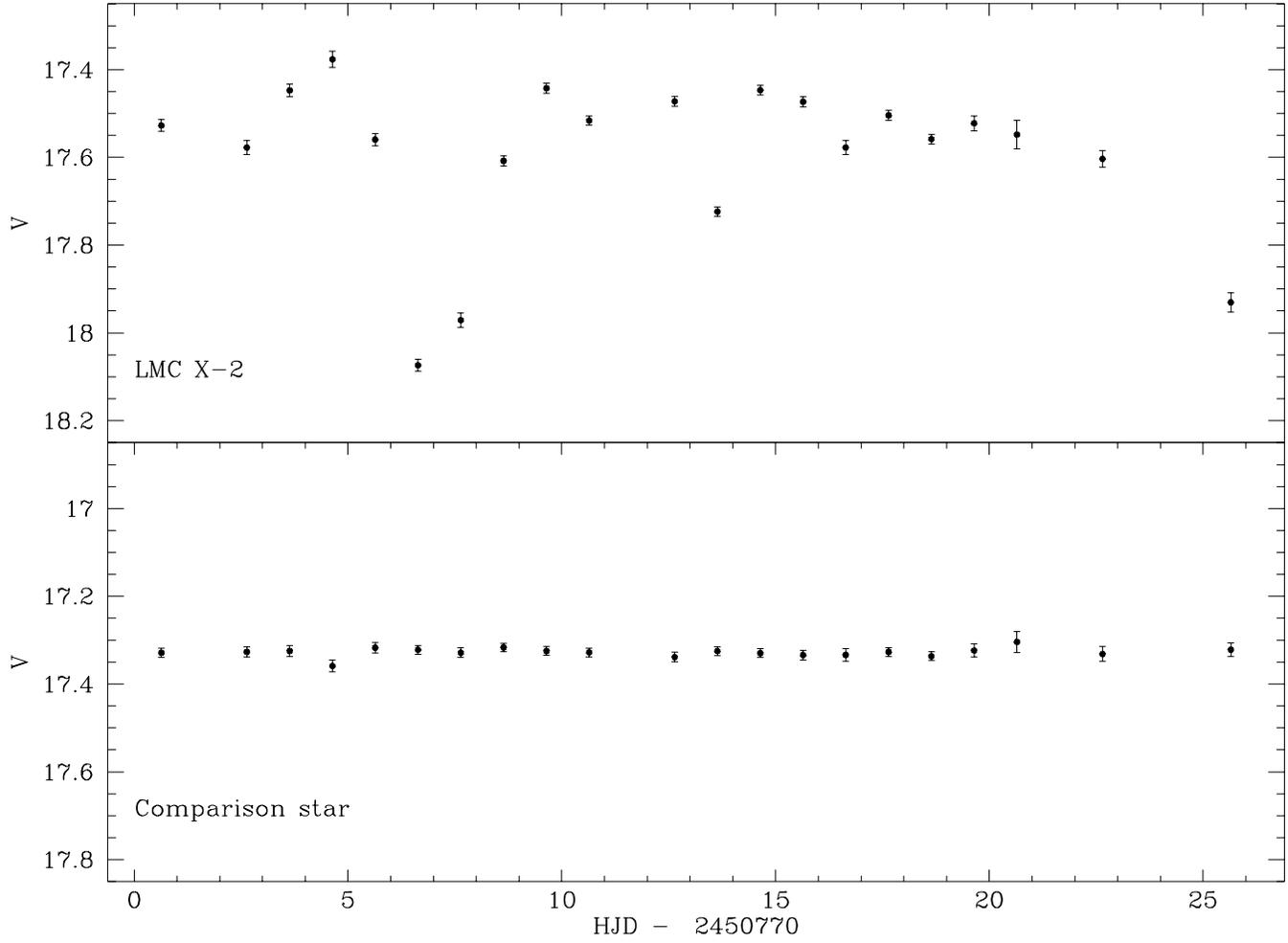}}}
\caption{ESO {\it V}-band light curve (300 s integrations) of LMC X-2 
({\it top}) and a star of comparable brightness ({\it bottom}).}
\end{figure*}

\label{lastpage}

\end{document}